\documentclass[reprint, amsmath,amssymb, aps]{revtex4-2}
\usepackage{graphicx}
\usepackage{dcolumn}
\usepackage{bm}
\usepackage{hyperref}
\usepackage{multirow}
\usepackage{ulem}

\begin{document}

\title{Spin-active single photon emitters in hexagonal boron nitride from carbon-based defects}

\author{Fernanda Pinilla}
\affiliation{Departamento de F\'isica, Facultad de Ciencias, Universidad de Chile. Santiago, Chile}
\author{Nicol\'as V\'asquez}%
\affiliation{Departamento de F\'isica, Facultad de Ciencias, Universidad de Chile. Santiago, Chile}
\author{Ignacio Chac\'on}%
\affiliation{Departamento de F\'isica, Facultad de Ciencias, Universidad de Chile. Santiago, Chile}
\author{Jerónimo R. Maze}
\affiliation{Instituto de Física, Pontificia Universidad Católica de Chile, Santiago, Chile}
\affiliation{Centro de Investigación en Nanotecnología y Materiales Avanzados (CIEN), Pontificia Universidad Católica de Chile, Santiago, Chile}
\author{Carlos C\'ardenas}
\affiliation{Center for the Development of Nanoscience and Nanotechnology (CEDENNA), Santiago, Chile}
\affiliation{Departamento de F\'isica, Facultad de Ciencias, Universidad de Chile. Santiago, Chile}
\author{Francisco Munoz}
\affiliation{Center for the Development of Nanoscience and Nanotechnology (CEDENNA), Santiago, Chile}
\affiliation{Departamento de F\'isica, Facultad de Ciencias, Universidad de Chile. Santiago, Chile}
 \email{fvmunoz@gmail.com}

\date{\today}

\begin{abstract}
Most single photon emitters in hexagonal boron nitride have been identified as carbon substitutional defects, forming donor-acceptor systems. Unlike the most studied bulk emitters (\textit{i.e.} color centers in diamond), these defects have no net spin, or have a single unpaired spin. By means of density functional calculations, we show that two non-adjacent carbon substitutional defects of the same type (\textit{i.e.} C$_\mathrm{B}$-C$_\mathrm{B}$, and C$_\mathrm{N}$-C$_\mathrm{N}$), can have a triplet groundstate. In particular, one of such defects has a zero phonon line energy of 2.5 eV, and its triplet state is nearly 0.5 eV more stable than its singlet. The mechanism behind the destabilization of the singlet state is related to a larger electrostatic repulsion of a symmetric wave function in a charged lattice.
\end{abstract}

\maketitle

\section{\label{sec:intro}Introduction}

The study and identification of the atomic defects associated with single photon emitters (SPEs) in hexagonal boron nitride (h-BN) is an active topic.\cite{White21,Rom_n_2021,Gale22,kianina2022}
Excluding  atomically engineered defects,\cite{Kianina20,Gao2021}  experiments  suggest C-based defects as SPEs.\cite{mendelson2021} So far, several substitutional C defects have been proposed. Those include monomers, different dimers (\textit{i.e.} including non-adjacent atoms),  trimers, arrangements of two close defects, and bigger substitutional clusters.\cite{jara2021,Auburger21,PhysRevMaterials.6.L042201,winter2021,PhysRevB.105.184101,huang2021,li2022,amblard2022universal}  All these defects follow a similar logic, when a C atom replaces a B(N) atom, \textit{i.e.} a C$_\mathrm{B}$ (C$_\mathrm{N}$) defect, it behaves as a donor(acceptor). Larger defects, with an even number of defect atoms, are built to have as many donors as acceptors, and the occupied and empty states will have a sizable band gap (we will elaborate on this at the beginning of Sec.~\ref{sec:results}). The case with an odd number of C atoms is similar, but with single uncompensated donor or acceptor, leading to a single extra energy level within the fundamental band gap. From this construction there exist only two possible spin configurations: (\textit{i}) if the number of C atoms of the defect is even, the SPE has no net spin; (\textit{ii}) if the number is odd, the SPE is paramagnetic with spin $S=\frac{1}{2}$.\cite{jara2021,Auburger21}

To the best of our knowledge, there are only a few types of C-based proposals of high-spin SPEs, with $S\ge 1$. One combines a substitutional and an out-of-plane interstitial defect.\cite{Bhang21} These defects offer an excellent example, beyond of the donor-acceptor logic, of a high-spin groundstate. The aforementioned defects have a set of zero phonon lines (ZPL) consistent with experimental results.\cite{Bhang21} Nevertheless, interstitials defects have a large formation energy and small migration barriers,\cite{Macia2022} and the natural occurrence of these defects should be rather scarce, due to their high formation energy. The second is a combination of a vacancy with a substitutional C atom.\cite{cheng2017} Again, the formation energy of these defects prevent them from being found in large concentrations.\cite{Macia2022} The formation energy, $H_f$ of a defect strongly depends on the chemical potential and Fermi energy, for single substitutional defects they are in the range $0<H_f<4.5$ eV. Excluding very specific combinations of Fermi energy and chemical potential, the $H_f$ of vacancies or interstitials is at least 2 eV higher than in substitutions.\cite{Macia2022,Bhang21,Weston2018}

The last proposal of SPEs with $S=1$, introduced by Maciaszek \textit{et al.}\cite{Macia2022}, is formed by four star-like substitutional C atoms: C$_\mathrm{B}(C_\mathrm{N})_3$ and C$_\mathrm{N}(C_\mathrm{B})_3$, they have a ZPL energy $\sim 2.0-2.2$ eV. These defects are an excellent example of pure substitutional C defects beyond from the standard donor-acceptor logic. However, these clusters only could be found in particular conditions (N-poor or N-rich), and with a rather low concentration (between two to three orders smaller than dimer or trimer defects).\cite{Macia2022}

Experimentally, optically-detected magnetic resonance (ODMR) has been employed to detect the spin from SPEs. Most of the measurements are associated with the V$_\mathrm{B}^-$ defect.\cite{Mathur2022,Baber2022,Yu2022} However, there is evidence of other spinful defects,\cite{Guo2021,mendelson2021,chejanovsky2021,Liu2022} with a photoluminescence spectra consistent with pure substitutional C defects. Particularly, a recent study\cite{Stern2022} has shown that a spin $S=1$ or $S=\frac{3}{2}$ is required to explain their ODMR measurements.

In this article, we show that two substitutional C defects of the same type (\textit{i.e.}  C$_\mathrm{N}$C$_\mathrm{N}$ or C$_\mathrm{B}$C$_\mathrm{B}$) can have a triplet groundstate, which in most cases is just a few meV lower in energy than the singlet. However, there is a specific defect where the triplet state is nearly 0.5 eV more stable. We  start by explaining our calculation methods in Sec.~\ref{sec:methods}. Then we  show our results (Sec.~\ref{sec:results}), starting with the simplest C$_\mathrm{B}$ and C$_\mathrm{N}$ defects and continuing with the dimer defects. Finally, in Sec.~\ref{sec:discuss}, we  provide an explanation of our findings.

\section{\label{sec:methods}Computational Methods}

Our central goal is to show that a triplet groundstate can be achieved with C-substitutional defects, even though it may seem counterintuitive at first glance. Then we will focus on the luminescent properties of the most interesting defect found. Our methodology reflects these goals, providing details about the accuracy of our calculations and measures to avoid possible methodological errors. Then, we explain the methodology used to calculate some properties: formation energy, ZPL energy, Huang-Rhys and Debye-Waller factors.

\subsection{Calculation parameters and approximations}
The calculations were performed with density functional theory (DFT), mostly by using the VASP package\cite{vasp1,vasp2,vasp3,vasp4}, but we also used the Gaussian code\cite{g09} in a few test calculations, as explained later in the text.

Regarding the settings of the VASP calculations (plane waves), the Perdew–Burke-Ernzerhof (PBE) and the Heyd–Scuseria–Ernzerhof (HSE) functionals were used.\cite{pbe, hse03,hse06} The PBE functional gives a  correct description of forces and geometries, however it has some shortcomings related to energies: (i) it underestimates the fundamental band gap of h-BN,\cite{Berseneva13} (ii) it underestimates the ZPL of C-based SPEs,\cite{jara2021} and (iii) it has problems representing the right distribution of localized states such as defect states. Conversely, the hybrid HSE functional gives quantitatively meaningful results, at a much higher computational cost.\cite{gali_hse} One of the most important advantages of the HSE functional is the inclusion of Hartree-Fock exchange in the short-range part of the Coulomb interaction, which is important for accurately describing the difference in energy between states of different multiplicity. The HSE functional has a free parameter, defining the limit between short- and long-range of the Coulomb interaction. We used the popular so-called HSE06 version of the functional, \textit{i.e.} the screening parameter is $\omega=0.11$ bohr$^{-1}$ and the relative amount of Hartree-Fock exchange was 1/4.\cite{hse06} Nevertheless, due to the decreased screening of 2D materials -such as h-BN- in some studies the value of the free parameter is changed to adjust the material's band gap.\cite{Auburger21} We do not expect a qualitative change by changing the form of the HSE functional (see Ref.~\cite{muechler2021quantum} for a longer discussion on this topic). 

Regarding the other parameters and settings of the calculations (VASP code), projector augmented-wave pseudopotentials were used.\cite{paw} A kinetic energy cutoff of 400 eV was employed, we tested a larger cutoff (650 eV)  and it did not change the results significantly (see the Appendix). A single k-point ($\Gamma$) was used in the supercell calculations. PyProcar was employed for the analysis of the results,\cite{pyprocar} and VESTA for the visualization.\cite{vesta} The geometries were relaxed with PBE in both the ground and excited states. {The relaxation with HSE06 of a specific defect, led to very similar results,\cite{Auburger21}  see Table~\ref{tab:convForces} in the Appendix. The lattice parameter was set to 2.49~\AA, the equilibrium value obtained with HSE06.}

An especially tricky point could be the usage of periodic boundary conditions (PBC), at least for the systems with a small difference between the singlet and triplet states. All the PBC calculations shown here use a $8\times 8$ supercell and the HSE06 functional. These calculations were tested at the PBE level with supercells ranging from $7\times 7$ to $10\times 10$. The differences due to the supercell size were minimal. Convergence tests of the supercell size with HSE06 are given in the Appendix. It is worth remarking that the HSE06 functional should be even less dependent on the supercell size, since it provides a much better localization of the defect's states.

To discard artifacts from the PBC, we used cluster models as explained in more detail in the Appendix. The shape of the clusters is critical for C$_\mathrm{N}$C$_\mathrm{N}$ clusters. Most cluster calculations were performed with VASP using the previous parameters, only the kinetic energy cutoff was increased to 500 eV, to better represent the finite size of the clusters. A few comparison calculations were performed with the Gaussian software, obtaining a good agreement between both codes. These calculations were done with the Gaussian 09.\cite{g09} For these calculations two double-Z basis set were used: 6-31+G(d) and cc-pVDZ.

\subsection{Calculation of physical properties}

The formation energy of a defect $D$ in a charge state $q$ is:
\begin{equation}
\begin{split}
    H_f(D,q) &= E_{tot}(D,q) - E_{hBN} - n_C\mu_C - n_B\mu_B\\ 
    &- n_N\mu_N + qE_F - E_{corr},    
\end{split}
\end{equation}
where $E_{tot}(D,q)$ is the total energy of a supercell hosting the defect in the desired charge state, $E_{hBN}$ is the energy of a pristine h-BN supercell, $\mu_i$ is the chemical potential of element $i$. $n_i$ is the change in the number of atoms of element $i$, with respect to the pristine supercell. The Fermi energy $E_F$ is set to zero at the valence band maximum. $E_{corr}$ are corrections of large importance in charged defects.\cite{RevModPhys.86.253, NAIK2018} Since the defects we studied are neutral, $q=0$, there is no need for an electrostatic correction. 

The chemical potential represents a reservoir for the exchange of atoms. We took as reference for $\mu_B$, $\mu_N$, and $\mu_C$, the energy per atom of bulk B, the N$_2$ molecule, and graphite, respectively. To simulate the limits of N-rich and N-poor conditions we followed the approach of Ref.~\cite{Weston2018}. 

The excited states were studied by means of the $\Delta$SCF method.\cite{PhysRevMaterials.5.084603} The singlet groundstate is spin degenerate and its excitation is from the last occupied level to the first empty. For the triplet groundstate, the excitation with lowest energy also is from the last occupied level to the first empty, but the spin channel depends on the system: in C$_\mathrm{B}$C$_\mathrm{B}$ the transition takes place on the majority spin, while in C$_\mathrm{N}$C$_\mathrm{N}$ the excitation takes place on the minority spin.

The Huang-Rhys factor $S$ measures the quantity of phonons participating in the ZPL transition. To obtain it, generalized coordinates $q_k$ are needed.\cite{Alkauskas_2014}
\begin{equation}
    q_k = \sum_{\alpha}\sqrt{m_{\alpha i}}(R_{e,\alpha i} - R_{g,\alpha i})\Delta r_{k,\alpha i},
    \label{g. coordinate}
\end{equation}
where the indexes $\alpha$, $i$ and $k$ denote atoms, Cartesian coordinates and vibrational modes, respectively. $R_e$ and $R_g$ are the positions in the excited and ground state, respectively. $m_\alpha$ is the mass of atom $\alpha$, and $\Delta r_k$ is the unitary vector of the $k$-th vibrational mode. Also, in this step we assumed that the vibrational modes for the ground and excited states are identical. In short, the above equation is a decomposition of the atomic rearrangement due to an optical excitation in the basis of the phonons.

The partial Huang-Rhys factor, $S_k$, quantifies the number of phonons of a given vibrational mode in the transition: 
\begin{equation}
    S_k = \frac{\omega_k q_k^2}{2\hbar},
    \label{Partial HR}
\end{equation}
where $\omega_k$ represents the frequency of the phonon. Finally, the Huang-Rhys factor is obtained by adding all these partial factors, $S=\sum_k S_k$. On the other hand, the Debye-Waller factor measures the weight of the ZPL peak, $w_{ZPL}$, with respect to the total photoluminescence spectrum. Since there is no emission of phonons at the ZPL, $w_{ZPL}=e^{-S}$.\cite{Alkauskas_2014}

\section{\label{sec:results}Results}

\subsection{\label{sec:cb}The C$_\mathrm{B}$ and C$_\mathrm{N}$ defects}

\begin{figure}[h]
    \centering
    \includegraphics[width=\columnwidth]{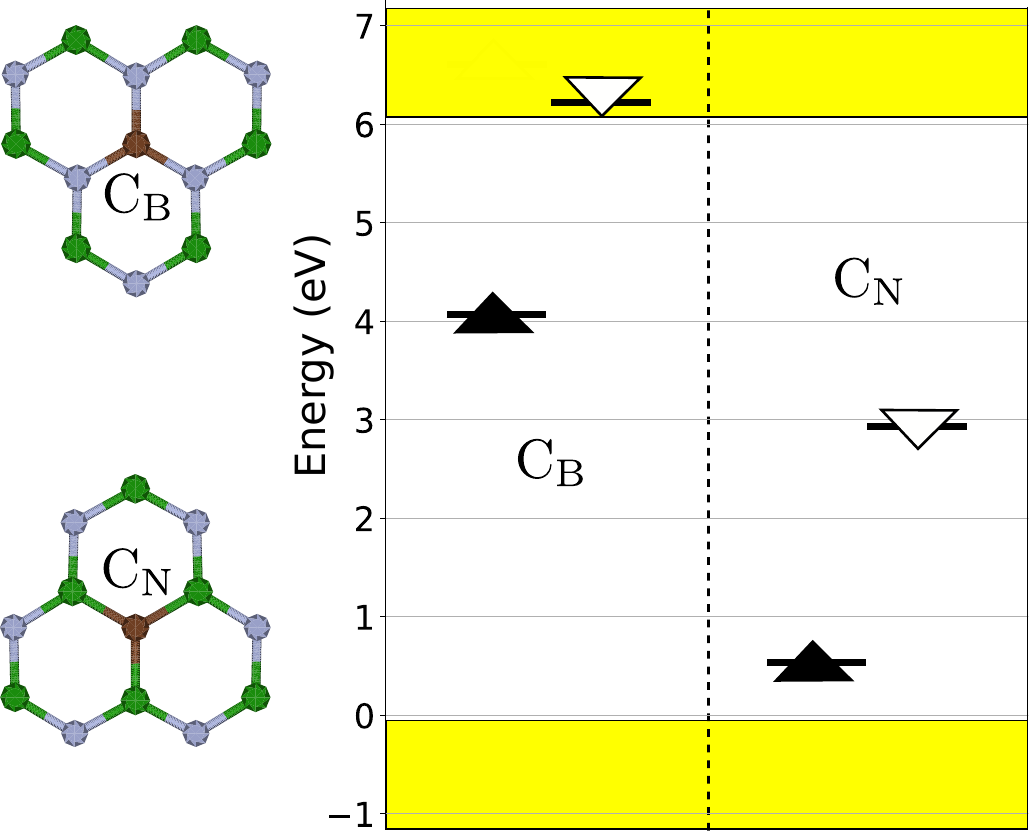}
    \caption{(left panel) Scheme of the C$_\mathrm{B}$ and C$_\mathrm{N}$ defects. (right panel) Representation of the relevant defect levels close to the fundamental band gap. The different spins are denoted by different symbols, and the occupied/empty bands are marked by filled/empty symbols. The valence and conduction bands are marked by a yellow region. The B, N , and C atoms are colored green, gray, and brown, respectively.}
    \label{fig:C1}
\end{figure}

The simplest substitutional C defects are the replacement of a B or N atom by a carbon, they are denoted C$_\mathbf{B}$ and C$_\mathrm{N}$, respectively, see Fig.~\ref{fig:C1}. When neutral, these defects are candidates for SPEs within the visible range.\cite{jara2021,Auburger21} In what follows we will focus only in the neutral cases. Both defects have groundstates within the fundamental band gap of h-BN. Each defect has a paramagnetic groundstate, $S=\frac{1}{2}$, and meanwhile C$_\mathrm{B}$ acts as a donor, with its last occupied level close to the conduction band, C$_\mathrm{N}$  acts like an acceptor with its first empty band near the middle of the band gap. If a C$_\mathrm{B}$ and a C$_\mathrm{N}$ are close enough, they form a dimer-like state, C$_\mathrm{B}^+$C$_\mathrm{N}^-$, which is non-magnetic. It shows luminescence with a ZPL that varies with the separation between the C$_\mathrm{N}$ and C$_\mathrm{B}$ defects.\cite{Auburger21}

\subsection{The C$_\mathrm{B}$C$_\mathrm{B}$ and C$_\mathrm{N}$C$_\mathrm{N}$ dimers}

\begin{figure}
    \centering
    \includegraphics[width=\columnwidth]{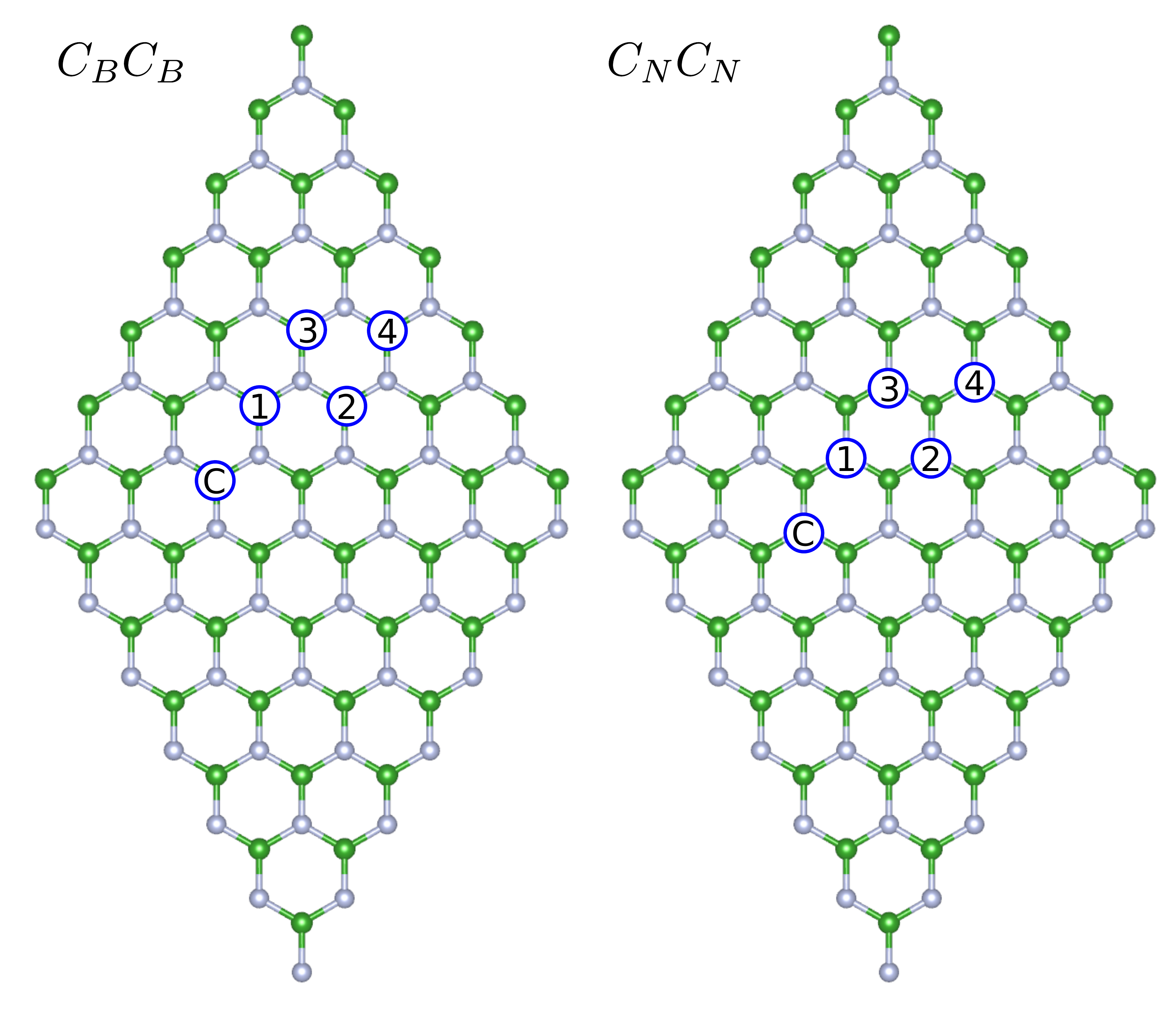}
    \caption{Conformations of a C$_\mathrm{B}$C$_\mathrm{B}$ and C$_\mathrm{N}$C$_\mathrm{N}$ dimer as a function of the distance, up to three lattice constants apart. One substitutional C atom is kept fixed and the other is labelled according to their distance, in ascending order. For instance the C$_\mathrm{N}$C$_\mathrm{N}$-3 dimer has one defect in the position marked `C', and the other in the position marked `3'.}
    \label{fig:geometry}
\end{figure}

The simplest way to avoid a donor-acceptor pair with a singlet groundstate is to consider two close defects of the same type, C$_\mathrm{B}$ or C$_\mathrm{N}$. Up to a few lattice sites, their localized states overlap, and they can be considered as a single defect. See Fig.~\ref{fig:geometry} for the geometry and labeling of the systems under study.

{
The formation energy, $H_f$, of the defects presented here is shown in Table \ref{tab:F.E} for two extreme conditions denoted as N-rich and N-poor, described in Sec.~\ref{sec:methods}. As expected, these values are similar to other C-based defects in h-BN, and their abundance should be determined by the actual environment.\cite{Macia2022}
\begin{table}[]
\caption{Formation energy $H_f$ of the different defects studied. They are reported for two extreme environments, N-poor and N-rich, as defined in Sec.~\ref{sec:methods}.}
\label{tab:F.E}
\begin{tabular}{|c|cc|}
\hline
\multirow{2}{*}{Configuration} & \multicolumn{2}{c|}{H$_f$ (eV)}       \\ \cline{2-3} 
                               & \multicolumn{1}{c|}{N-poor} & N-rich \\ \hline
C$_\mathrm{N}$C$_\mathrm{N}$-2                     & \multicolumn{1}{c|}{3.82}   & 9.02   \\ \hline
C$_\mathrm{N}$C$_\mathrm{N}$-3                     & \multicolumn{1}{c|}{3.82}   & 9.02   \\ \hline
C$_\mathrm{N}$C$_\mathrm{N}$-4                     & \multicolumn{1}{c|}{3.83}   & 9.03   \\ \hline
C$_\mathrm{B}$C$_\mathrm{B}$-2                     & \multicolumn{1}{c|}{8.47}   & 3.27   \\ \hline
C$_\mathrm{B}$C$_\mathrm{B}$-3                     & \multicolumn{1}{c|}{8.49}   & 3.29   \\ \hline
C$_\mathrm{B}$C$_\mathrm{B}$-4                     & \multicolumn{1}{c|}{8.47}   & 3.27   \\ \hline
\end{tabular}

\end{table}
}

\begin{figure}
    \centering
    \includegraphics[width=0.8\columnwidth]{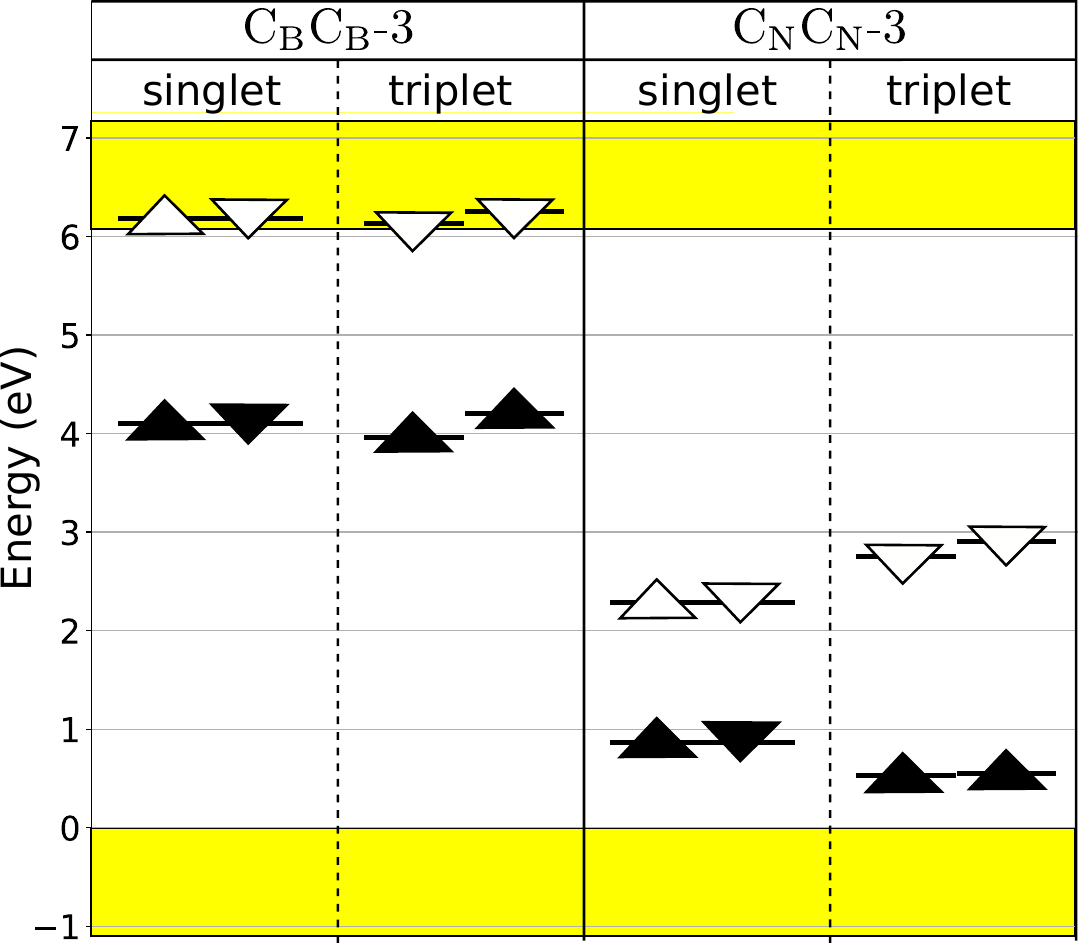}
    \caption{Energy levels of the C$_\mathrm{B}$C$_\mathrm{B}$-3 and C$_\mathrm{N}$C$_\mathrm{N}$-3 defects, for the singlet and triplet states. The symbols are the same used in Fig.~\ref{fig:C1}. The distance between both atoms is nearly 5~\AA, see Table~\ref{tab:summary}.}
    \label{fig:levelsDimer}
\end{figure}

\begin{figure*}[t]
    \centering
    \includegraphics[width=0.75\textwidth]{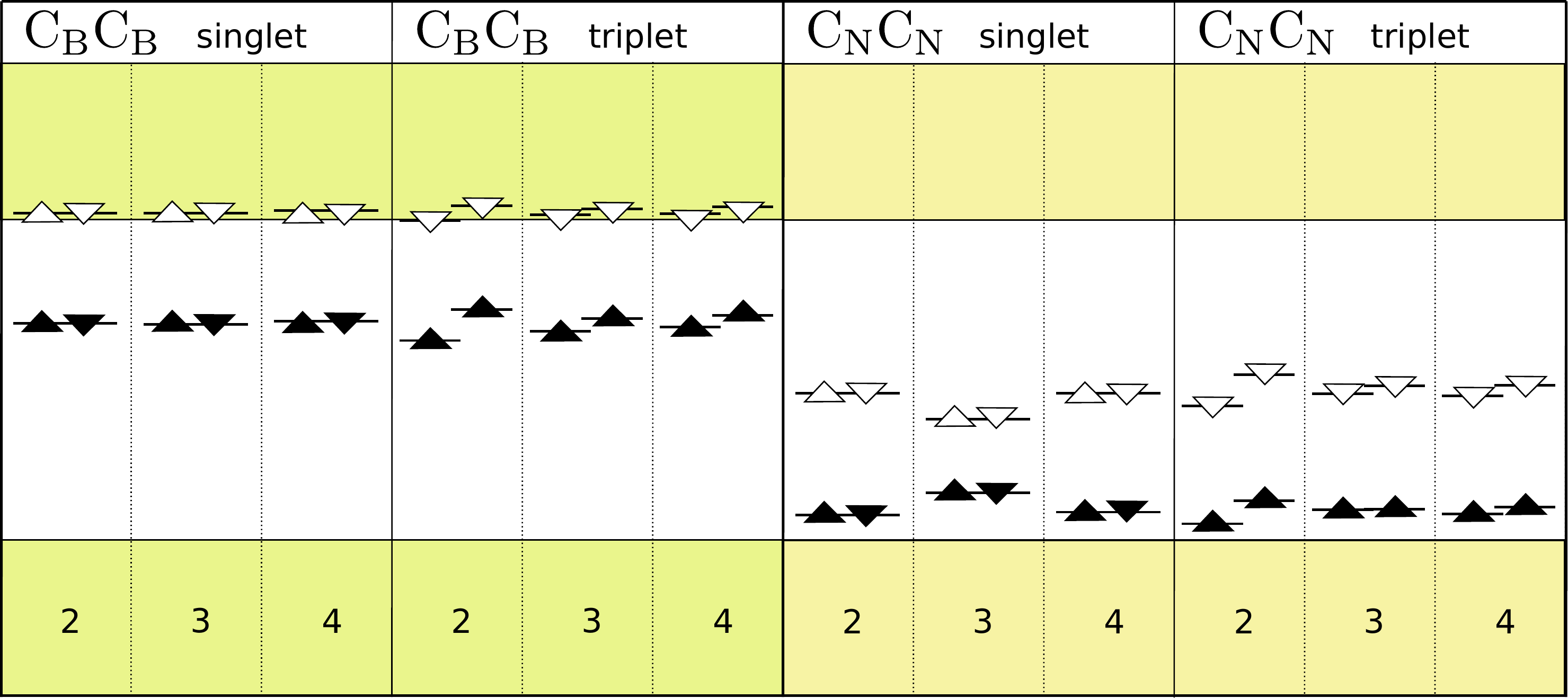}
    \caption{Energy levels of the C$_\mathrm{B}$C$_\mathrm{B}$-X and C$_\mathrm{N}$C$_\mathrm{N}$-X defects, the distance is specified in the lower part of each panel. The symbols are the same used in Fig.~\ref{fig:C1}. The different color on the left and right sides is just a guide to the eye.}
    \label{fig:c2-all}
\end{figure*}

When both defect atoms are not too close (\textit{i.e.} beyond nearest neighbors), their levels can be inferred up to some extent from the monomers, see Fig.~\ref{fig:C1} as reference, with two possible arrangements, the singlet and the triplet. The actual energy levels  for a specific distance are shown in Fig.~\ref{fig:levelsDimer}. The visual inspection of the C$_\mathrm{B}$C$_\mathrm{B}$-3 dimer shows no appreciable difference between the single and triplet arrangements. However, C$_\mathrm{N}$C$_\mathrm{N}$-3 shows a much larger separation between the occupied and empty defect levels in the triplet, lowering the total energy.
{ The different values of the defect levels as a function of the distance is shown in Fig.~\ref{fig:c2-all}. The defects of the C$_\mathrm{B}$C$_\mathrm{B}$ family do not show any appreciable difference in the singlet states as a function of the distance. However, in the triplet there is an appreciable distance-dependent interaction for the same spin value. Just from inspecting the energy levels the groundstate is not evident, \textit{e.g.} the singlet and triplet have a comparable energy. The C$_\mathrm{N}$C$_\mathrm{N}$ defects show a similar pattern with the exception of C$_\mathrm{N}$C$_\mathrm{N}$-3, which was already discussed and it will be explained in more detail in Sec.~\ref{sec:discuss}.}

\begin{table}[h]
    \centering
        \caption{Energy difference $\Delta E^{(S-T)}$ = $E_{singlet} - E_{triplet}$ for the studied cluster (\textit{i.e.} a negative value implies a singlet groundstate). The values are obtained with the HSE06 functional, and using the PBE positions. The distances correspond to the singlet state, the values between parentheses correspond to the triplet state. }
    \label{tab:summary}

\begin{tabular}{|l|r|r|r|}
    \hline
         Label & distance (\AA) & $\Delta E^{(S-T)}$ (meV) \\\hline
         C$_\mathrm{B}$-C$_\mathrm{B}$-2 & 4.32 (4.32) & -27  \\\hline
         C$_\mathrm{B}$-C$_\mathrm{B}$-3 & 4.98 (4.98) &  32 \\\hline
         C$_\mathrm{B}$-C$_\mathrm{B}$-4 & 6.61 (6.61) &  20 \\\hline\hline
         C$_\mathrm{N}$-C$_\mathrm{N}$-2 & 4.33 (4.35) & -44 \\\hline
         C$_\mathrm{N}$-C$_\mathrm{N}$-3 & 5.02 (5.03) & 482 \\\hline
         C$_\mathrm{N}$-C$_\mathrm{N}$-4 & 6.61 (6.62) &  19 \\\hline
    \end{tabular}
\end{table}

The actual differences in total energy of all the systems studied are in Table~\ref{tab:summary}. In general, for closer defects (C$_X$C$_X$-2) the singlet state is slightly more stable. For longer distances (C$_X$C$_X$-4) the opposite is true. In the remaining case, C$_X$C$_X$-3, for B defects, shows an intermediate behavior, with both configurations almost degenerate in energy. However, in C$_\mathrm{N}$C$_\mathrm{N}$-3 the triplet is nearly 0.5 eV more stable than the singlet. Although this last result may seem strange at first glance, in the next section we will show that a lower energy for the triplet is to be expected. Additionally, an extensive set of convergence tests, including cluster models, is in the Appendix.

The defects C$_\mathrm{X}$C$_\mathrm{X}$-3 share have in common a triplet groundstate. They can have an optical transition from the valence band to the unoccupied defect states. The zero phonon line (ZPL) of C$_\mathrm{N}$C$_\mathrm{N}$-3 has energy $E_{ZPL}=2.5$ eV, if the lowest energy level is involved in the transition. { Its Huang-Rhys and Debye-Waller factors are $S = 1.37$ and $w_{ZPL}=0.25$, respectively. In contrast, the defect C$_\mathrm{B}$C$_\mathrm{B}$-3, has a ZPL energy $E_{ZPL}=1.6$ eV, a much smaller electron-phonon, with $S=0.67$ and $w_{ZPL}=0.51$. 

For the sake of comparison, experiments in h-BN, have found a myriad of SPEs, with wavelength covering the visible spectrum,\cite{Wigger_2019} likely most of them are C-substitutional defects forming a donor-acceptor pair. Experimental values of the Huang-Rhys factor\cite{PhysRevMaterials.6.L042201} are $S=1.19\pm0.43$ ($w_{ZPL}=0.33\pm 0.13$). Also it is insightful to compare  C$_\mathrm{X}$C$_\mathrm{X}$-3 defects with the NV$^-$ and SiV$^0$ defects found in bulk diamond, all of them with a triplet groundstate. The ZPL energy of the NV$^-$ center is $E_{ZPL}=1.94$ eV, and a very prominent phonon side-band, $S=3.49$ ($w_{ZPL}=0.03$).\cite{Kurt2000} The SiV$^0$ center has $E_{ZPL}=1.67$ eV, and a remarkably small electron-phonon coupling, $S=0.24$ ($w_{ZPL}=0.79$).\cite{haussler2017} In this respect, h-BN C-based defects have some clear advantages and drawbacks compared with their bulk diamond  counterparts. First, a 2D material hosting the SPEs allows a much easier integration into devices, even more, there is unprecedented possibility of controlling the SPE by forming a van der Waals heterostructure.\cite{caldwell2019} Second, C-based SPEs in h-BN offer a `ultrabright' stable luminescent signal, overcoming the signal of the NV$^-$ and SiV centers.\cite{tran2016quantum} We expect a similar behavior of the SPEs presented here. Third, the SPEs studied here have a small electron-phonon coupling, favoring long coherence times. However, the SiV center has an even smaller electron-phonon coupling, useful for quantum technologies. Fourth, there is a widespread problem to generate specific C-based SPEs in h-BN. We are aware of two reports of monochromatic SPEs in h-BN, blue\cite{Gale22} and yellow,\cite{kumar2023localized} neither of them seems to be spin-active. 

}

The energy difference $\Delta E^{(S-T)}$ of most  C$_\mathrm{X}$C$_\mathrm{X}$ defects is smaller than room temperature ($\sim 26$ meV), but larger than the critical temperature of liquid nitrogen (7 meV), which is easy to reach in experiments and make our predictions susceptible to experimental test. The triplet state in the defect C$_\mathrm{N}$C$_\mathrm{N}$-3 should remain stable even near the dissociation temperature of h-BN ($\sim 3000$ K).

\section{\label{sec:discuss}Discussion}

\begin{figure}
    \centering
    \includegraphics[width=0.7\columnwidth]{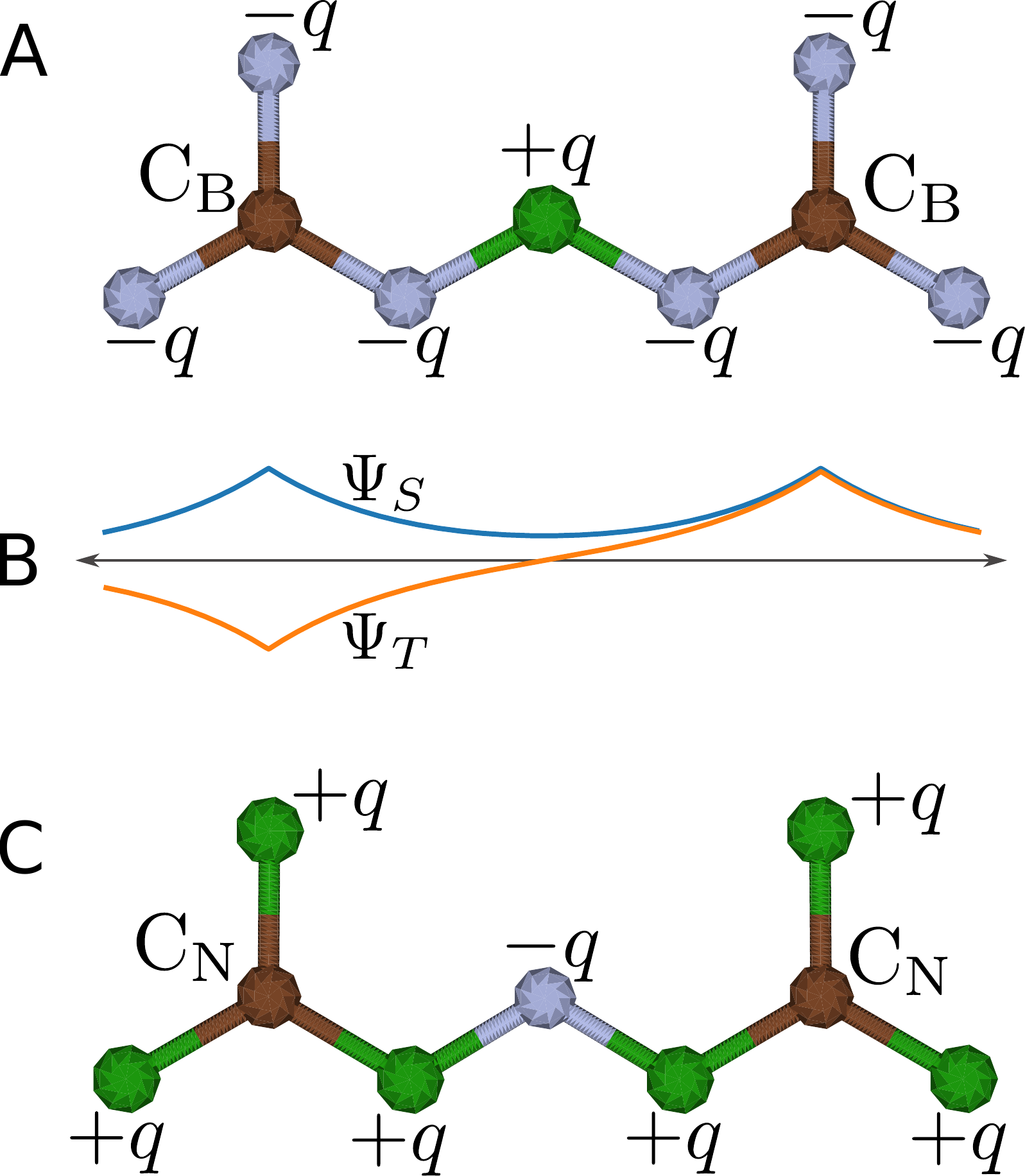}
    \caption{Atomic environment for the C$_\mathrm{B}$C$_\mathrm{B}$-3 (panel A) and C$_\mathrm{N}$C$_\mathrm{N}$-3 (panel C) clusters. The charges centered at each atom are denoted by $\pm q$. The central panel (B) is a sketch of a typical singlet ($\Psi_S$, nodeless) and triplet ($\Psi_T$, with a node at its middle point) wave functions.} 
    \label{fig:zoom}
\end{figure}

A simple model for the C$_\mathrm{X}$C$_\mathrm{X}$ defects is a two-electron molecule in an effective medium, similar to the H$_2$ molecule, but with extra terms in the Hamiltonian due to the h-BN lattice. The Hamiltonian is:
\begin{equation}
    H = H_0 + \Delta V,
\end{equation}
\noindent where $H_0$ is the standard Hamiltonian of a H$_2$-like system,
\begin{equation}
\begin{split}
    H_0 &= -\frac{\nabla^2_1}{2} - \frac{\nabla^2_2}{2}
    - \frac{1}{r_{1A}} - \frac{1}{r_{2B}} - \frac{1}{r_{1B}}\\
    &- \frac{1}{r_{2A}}
    + \frac{1}{r_{12}} + \frac{1}{R_{AB}}    
\end{split}
    \label{eq:hamiltonian1}
\end{equation}
\noindent with the sub-indexes \{1,2\} labelling both electrons, and the substitutional ions are denoted as \{A,B\}. The distances are $r_{\alpha\beta}=|r_\alpha-r_\beta|$. The contribution of the lattice to the C$_\mathrm{X}$C$_\mathrm{X}$ Hamiltonian can be modeled as an electrostatic-only contribution:
\begin{equation}
    \Delta V = -\sum_{i\neq\{A,B\}}\frac{q_i}{r_{1i}} - \sum_{i\neq\{A,B\}}\frac{q_i}{r_{2i}}.
\end{equation}
Here,  $q_i$ the effective charge of each site, which is negative (positive) for the N (B) atoms, see Fig.~\ref{fig:zoom}.

The solutions to Eq. \ref{eq:hamiltonian1} can be symmetric in the orbitals or in the spin, forming a singlet or a triplet, respectively. Without the lattice potential ($\Delta V=0$), the relative energy of both states is given by  \textit{exchange} and \textit{Coulomb} integrals. The singlet state always has a lower energy, as it is well established \cite{heitler1927}.

\begin{figure}
    \centering
    \includegraphics[width=0.7\columnwidth]{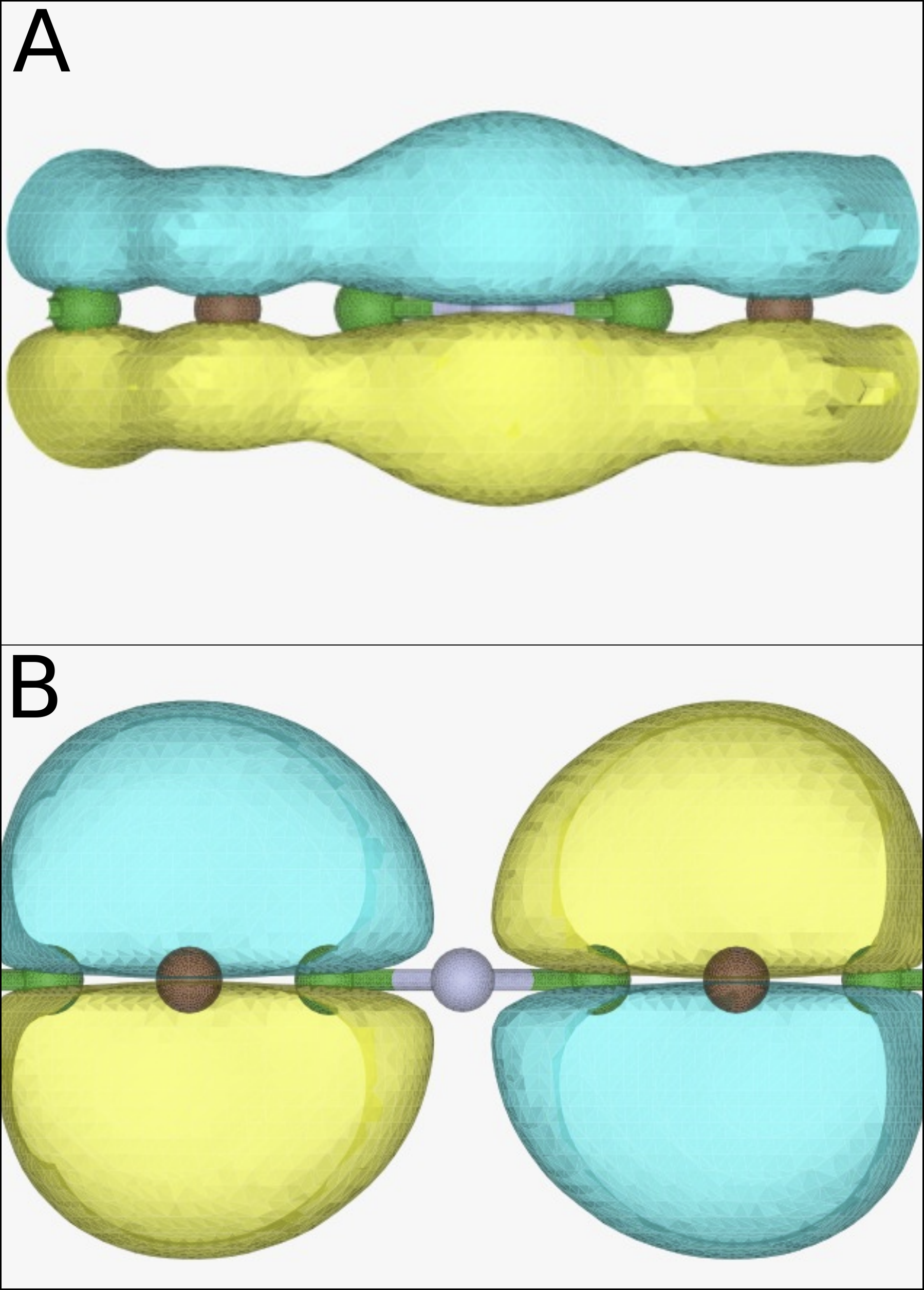}
    \caption{Wave functions of one of the occupied defect states of the C$_\mathrm{N}$C$_\mathrm{N}$-3 defect for the singlet (panel A) and triplet (panel B) states. The C-N-C atoms are shown. The wave functions corresponds to a PBE-level calculation.}
    \label{fig:wf}
\end{figure}

Considering  exponentially decaying states for the defect wave functions, the energy associated to $\Delta V$  will be similar for both the singlet and triplet for most  lattice sites. This is because the localized wave function is relevant only for a few nearest neighbors, and the effects of having a (anti-) symmetric wave function are relevant only where the contribution of both states is non-negligible. However, there is a position of the lattice which is particularly relevant, at least for the C$_\mathrm{X}$C$_\mathrm{X}$-3 defects. See Fig.~\ref{fig:zoom}, the atom at the midpoint between both C atoms is located in a region where the amplitude of the wavefunctions of the defect is not negligible.  This atom is only  $\sim$2.5~\AA~  apart from the C positions. Also, it can have a positive or negative net charge, which implies that this site  is attractive for electrons in the case of the  C$_\mathrm{B}$C$_\mathrm{B}$-3 defect and repulsive in the case of C$_\mathrm{N}$C$_\mathrm{N}$-3. If we label this atoms as $M$, the energy associated to this site is 

\begin{equation}
    V_M^{\alpha}\equiv\left\langle\Psi_\alpha\left| - \frac{q_M}{r_{1M}} - \frac{q_M}{r_{2M}} \right| \Psi_\alpha \right\rangle, 
\end{equation}

where $\alpha=\{S,T\}$ for a singlet or triplet state. For C$_\mathrm{B}$C$_\mathrm{B}$-3,  since the charge $q_M$ is positive, a larger amplitude of $|\Psi_\alpha\rangle$  in this atom is favored. This is achieved in the nodeless singlet state, and $V_M^{\alpha}$ is such that $V^S_M<V^T_M<0$. Contrary , for C$_\mathrm{N}$C$_\mathrm{N}$-3, the potential due to the site $M$ is repulsive for electrons, and the anti-symmetry of the triplet lowers the total energy: $0<V^T_M<V^S_M$. Therefore, $\Delta V$ favors a triplet (singlet) for C$_\mathrm{N}$C$_\mathrm{N}$-3 (C$_\mathrm{B}$C$_\mathrm{B}$-3). The difference between the singlet and triplet wave function for C$_\mathrm{N}$C$_\mathrm{N}$-3 is shown in Fig.~\ref{fig:wf}, note the absence or presence of a central node. It is worth mentioning that with this simple analysis we only intend to provide an \textit{ex post} explanation of our results. The simplicity of the model does not allow us to predict which will be the most stable state, the singlet or the triplet.

\section{\label{sec:conclusions}Conclusions}
By means of DFT calculations we found that two substitutional C$_\mathrm{N}$ or C$_\mathrm{B}$ defects of the same type, have a triplet ($S=1$) groundstate when both C atoms are sufficiently separated. In particular, when two C$_\mathrm{N}$ atoms are separated by two lattice parameters ($\approx 5$~\AA, denoted C$_\mathrm{N}$C$_\mathrm{N}$-3 in the article) the triplet state is lower in energy by $\approx 0.5$ eV, compared with the singlet state. The ZPL associated with this defect is $E_{ZPL}=2.53$ eV. In the remaining cases the energy difference between the triplet and singlet states is smaller than $0.05$ eV. 

The defect C$_\mathrm{N}$C$_\mathrm{N}$-3 could explain the measured defects in h-BN with $S>\frac{1}{2}$.\cite{Guo2021,mendelson2021,Liu2022,Stern2022}
 Since no vacancy is involved, they should have a large abundance\cite{Macia2022}. Also, as pure substitutional C defects, they should exhibit the typical phonon sideband (PSB) with lateral maximum shift $~160-180$ meV from the ZPL.\cite{jara2021}.

On the other hand, the remaining non-adjacent C$_\mathrm{N}$C$_\mathrm{N}$ and C$_\mathrm{B}$C$_\mathrm{B}$ defects have a rather small energy difference between both the triplet and singlet state. Therefore, it could be possible to induce a switch between both states by means of an external field, or in a van der Waals heterostructure.

\section{Competing Interests}
The Authors declare no Competing Financial or Non-Financial Interests

\section{Data Availability}
All input files are available from the corresponding author upon reasonable request.

\section{Author Contributions}
F.P., C.C., N.V and F.M. run all the calculations. F.M. conceived the idea. F.M., J.R.M., and C.C. explained the stability of the triplet state. All authors discussed the results and contributed to the writing of the manuscript.

\begin{acknowledgments}
This work was partially supported by Fondecyt Grants No. 1191353, 1220715, 1220366, and 1221512 by the Center for the Development of Nanoscience and Nanotechnology CEDENNA AFB180001, and from Conicyt PIA/Anillo ACT192023. This research was partially supported by the supercomputing infrastructure of the NLHPC (ECM-02).
\end{acknowledgments}

\appendix

\section{Convergence and boundary conditions}

The results from Table~\ref{tab:summary}, shows a drastic difference between one particular defect and the others. It is valid to ask whether these results are an artifact from the calculations or not. In this aspect we will show that for a large set of parameters the results do hold.

The first aspect to test is the role of the supercell size. We focus on the trends of the C$_\mathrm{N}$-C$_\mathrm{N}$ defects with the HSE06 functional, since these defects show the abrupt changes. Table~\ref{tab:convSupercell} shows the same behavior regardless of the size of the supercell. For the most separated defects, C$_\mathrm{N}$-C$_\mathrm{N}$-4, the supercell influences the results in the range of a few meV. 

\begin{table}[h]
    \caption{Convergence respect to supercell size. The other parameters are given in Sec.~\ref{sec:methods}.}
    \label{tab:convSupercell}
    \centering
    \begin{tabular}{|c|c|r|}
    \hline
    Label         & Supercell & $\Delta E^{(S-T)}$ (meV)\\\hline
    C$_\mathrm{N}$-C$_\mathrm{N}$-2 & $6\times 6$ & -44         \\\hline
    C$_\mathrm{N}$-C$_\mathrm{N}$-2 & $7\times 7$ & -44         \\\hline
    C$_\mathrm{N}$-C$_\mathrm{N}$-2 & $8\times 8$ & -44         \\\hline\hline
    C$_\mathrm{N}$-C$_\mathrm{N}$-3 & $6\times 6$ & 486         \\\hline
    C$_\mathrm{N}$-C$_\mathrm{N}$-3 & $7\times 7$ & 482         \\\hline
    C$_\mathrm{N}$-C$_\mathrm{N}$-3 & $8\times 8$ & 482         \\\hline\hline
    C$_\mathrm{N}$-C$_\mathrm{N}$-4 & $6\times 6$ &  27         \\\hline
    C$_\mathrm{N}$-C$_\mathrm{N}$-4 & $7\times 7$ &  15         \\\hline
    C$_\mathrm{N}$-C$_\mathrm{N}$-4 & $8\times 8$ &  19         \\\hline
    \end{tabular}
\end{table}

The completeness of the plane-wave basis is given by the kinetic energy cutoff, and in defect calculations it is critical to resolve localized states. We extensively tested  the cutoff with the PBE functional (not shown here). Obtaining that a cutoff of 400 eV is good enough to have an accuracy of a few meVs. However, our results suggest that PBE is unable to correctly capture the localization of the defect's electron density. To check the accuracy of our results with a cutoff of 400 eV, we tested a cutoff of 650 eV for two defects with the most interesting/complex interplay between localization and electronic occupations. While at first glance it may be surprising that a cutoff of 400 eV is enough to describe localized defect states, this is the cutoff recommended in the pseudopotential to describe localized states, such as the h-BN bonds.

\begin{table}[h]
    \caption{Effect of the kinetic energy cutoff. The other parameters are given in Sec.~\ref{sec:methods}.}
    \label{tab:convCutoff}
    \centering
    \begin{tabular}{|c|c|r|}
    \hline
    Label         & cutoff (eV) & $\Delta E^{(S-T)}$ (meV)\\\hline
    C$_\mathrm{N}$-C$_\mathrm{N}$-3 & 400 & 482         \\\hline
    C$_\mathrm{N}$-C$_\mathrm{N}$-3 & 650 & 484         \\\hline
    C$_\mathrm{B}$-C$_\mathrm{B}$-3 & 400 & 32         \\\hline
    C$_\mathrm{B}$-C$_\mathrm{B}$-3 & 650 & 31         \\\hline
    \end{tabular}
\end{table}

Another test is the effect of the atomic relaxation with HSE06, compared with the relaxation with PBE followed by a static calculation with HSE06. Again, we only tested the C$_\mathrm{N}$-C$_\mathrm{N}$ defects. Table~\ref{tab:convForces} shows a minor effect of the functional regarding the relative stability of the singlet/triplet states. This is because \textit{(i)} PBE provides decent forces and positions, and \textit{(ii)} the error in positions due to PBE is likely to be consistent (\textit{e.g.} longer C-N bonds) and cancelled when measuring relative energies.

\begin{table}[h]
    \caption{Effect of using the positions relaxed with PBE plus a HSE06 calculation without relaxation, compared to a full HSE06 relaxation. In both cases the results are calculated with a $7\times 7$ supercell. The remaining parameters are given in Sec.~\ref{sec:methods}}
    \label{tab:convForces}
    \centering
    \begin{tabular}{|c|c|r|}
    \hline
    Label         & relaxation & $\Delta E^{(S-T)}$ (meV) \\\hline
    C$_\mathrm{N}$-C$_\mathrm{N}$-2 & HSE06        & -44      \\\hline
    C$_\mathrm{N}$-C$_\mathrm{N}$-2 & PBE          & -44      \\\hline
    C$_\mathrm{N}$-C$_\mathrm{N}$-3 & HSE06        & 474      \\\hline
    C$_\mathrm{N}$-C$_\mathrm{N}$-3 & PBE          & 482      \\\hline
    C$_\mathrm{N}$-C$_\mathrm{N}$-4 & HSE06        &  16      \\\hline
    C$_\mathrm{N}$-C$_\mathrm{N}$-4 & PBE          &  15      \\\hline     
    \end{tabular}
\end{table}

\begin{figure}
    \centering
    \includegraphics[width=0.9\columnwidth]{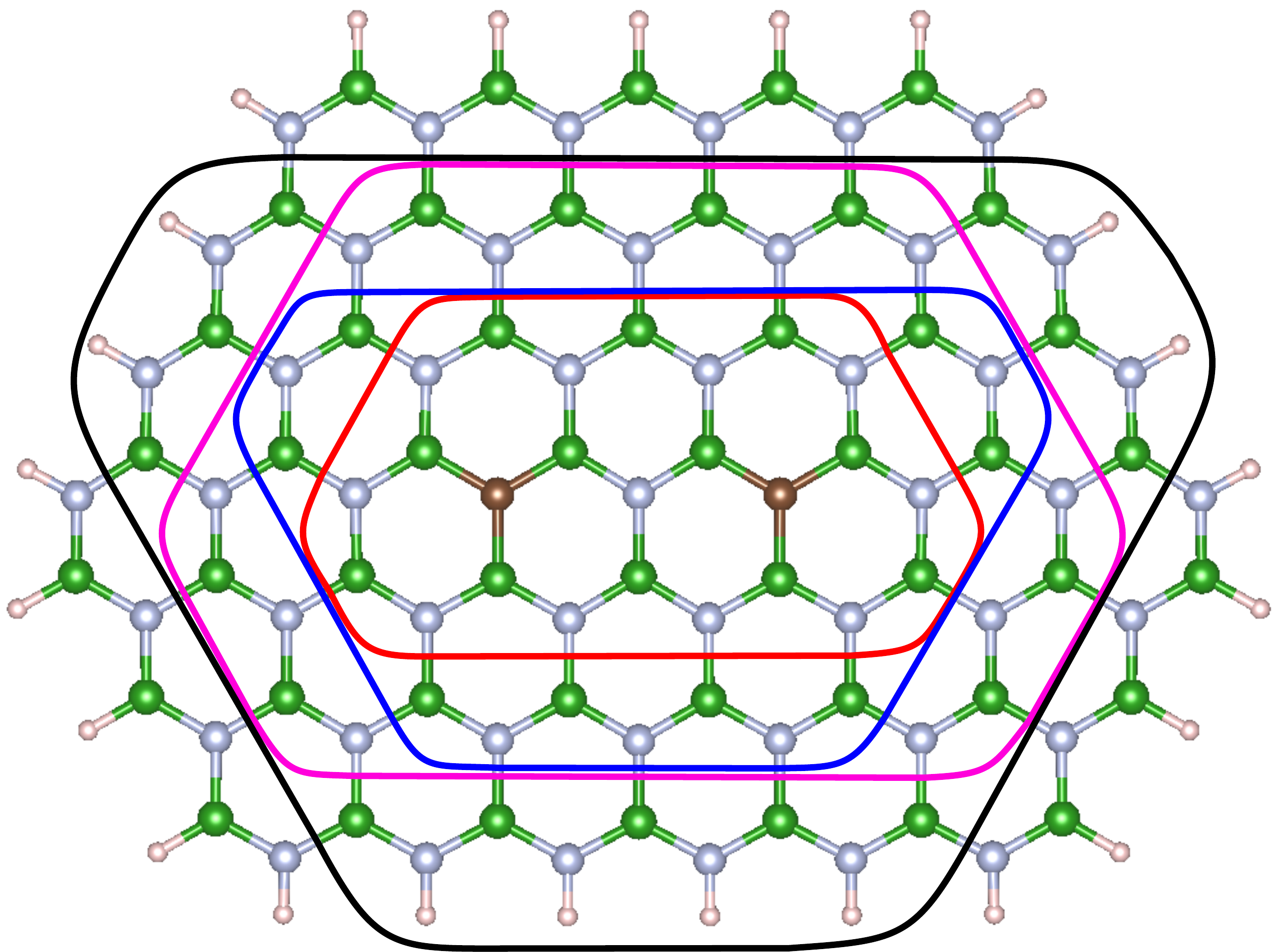}
    \caption{Cluster models used in this test. The size (ignoring H atoms) of the clusters used is marked in the figure. The clusters are B$_{12}$N$_{11}$C$_{2}$ (inside red region), B$_{20}$N$_{16}$C$_{2}$ (inside blue region), B$_{28}$N$_{27}$C$_{2}$ (inside magenta region), B$_{39}$N$_{35}$C$_{2}$ (inside black region), B$_{50}$N$_{49}$C$_{2}$ (the full cluster in the figure). The smaller white atoms are H.}
    \label{fig:cluster}
\end{figure}

The last test we made, is to use a cluster model to rule out artifacts from the PBC. The clusters are shown in Fig.~\ref{fig:cluster}. All the clusters are passivated with H atoms, and were created with pyPoscar\cite{pyposcar}. The results are given in Table~\ref{tab:clusters}, and show a good agreement with supercell calculations. { A strange trend is evident from the Table~\ref{tab:clusters} with a minimum energy $\Delta E^{(S-T)}$ at B$_{28}$N$_{27}$C$_{2}$, and it increases for larger and smaller clusters. We interpret such behavior in terms of the localization of the defect wave functions, for smaller clusters it is spuriously localized, resulting in an increased interaction for both spin configurations. As the size of the system increases, the cluster boundaries become less relevant. However, it does not happen at the same rate. The triplet state is less localized, needing a larger cluster to avoid artifacts due to the finite boundaries. That extra interaction results in lower total energy of the triplet, in concordance with Table~\ref{tab:clusters}. Since we used a larger cutoff energy and supercell size in these calculations, a perfect agreement with the results of periodic systems is unlikely.}

\begin{table}[h]
    \caption{Cluster models for the defect C$_\mathrm{N}$C$_\mathrm{N}$-3. The shape of the cluster is shown in Fig.~\ref{fig:cluster}. The kinetic energy cutoff was set to 500 eV, the remaining parameters are given in Sec.~\ref{sec:methods}.}
    \label{tab:clusters}
    \centering
    \begin{tabular}{|c|c|r|}
    \hline
    Cluster                 &  $\Delta E^{(S-T)}$ (meV) \\\hline
    B$_{12}$N$_{11}$C$_{2}$ &  522     \\\hline
    B$_{20}$N$_{16}$C$_{2}$ &  500     \\\hline
    B$_{28}$N$_{27}$C$_{2}$ &  487     \\\hline
    B$_{39}$N$_{35}$C$_{2}$ &  489     \\\hline
    B$_{50}$N$_{49}$C$_{2}$ &  495     \\\hline
    \end{tabular}
\end{table}

\bibliography{bib}% Produces the bibliography via BibTeX.

\end{document}